\documentclass[%
 reprint,
 amsmath,amssymb,
 aps,
]{revtex4-2}
\usepackage[inkscapelatex=false]{svg}
\usepackage{graphicx}
\usepackage{dcolumn}
\usepackage{bm}
\usepackage{todonotes}
\newtheorem{definition}{Definition}
\usepackage{hyperref}

\begin{document}

\title{Machine Learning Message-Passing for the Scalable Decoding of QLDPC Codes}
\author{Arshpreet Singh Maan}
    \email{arshpreet.maan@aalto.fi}
\author{Alexandru Paler}%
 \email{alexandru.paler@aalto.fi}
\affiliation{%
 Aalto University, Espoo, Finland\\
}%

\begin{abstract}
We present Astra, a novel and scalable decoder using graph neural networks. Our decoder works similarly to solving a Sudoku puzzle of constraints represented by the Tanner graph. In general, Quantum Low Density Parity Check (QLDPC) decoding is based on Belief Propagation (BP, a variant of message-passing) and requires time intensive post-processing methods such as Ordered Statistics Decoding (OSD). Without using any post-processing, Astra achieves higher thresholds and better logical error rates when compared to BP+OSD, both for surface codes trained up to distance 11 and Bivariate Bicycle (BB) codes trained up to distance 18. Moreover, we can successfully extrapolate the decoding functionality: we decode high distances (surface code up to distance 25 and BB code up to distance 34) by using decoders trained on lower distances. Astra+OSD is faster than BP+OSD. We show that with decreasing physical error rates, Astra+OSD makes progressively fewer calls to OSD when compared to BP+OSD, even in the context of extrapolated decoding. Astra(+OSD) achieves orders of magnitude lower logical error rates for BB codes compared to BP(+OSD). The source code is open-sourced at \url{https://github.com/arshpreetmaan/astra}.
\end{abstract}

\maketitle

\section{Introduction}

The successful decoding of Quantum Error Correction (QEC) codes plays a significant role in achieving large scale, practical quantum computations. The accuracy (i.e. reduce the probability of logical error) and scalability (i.e. increasing the QEC code distance should not significantly slow down the decoding process) are some of the most important characteristics of a decoder.

There are various QEC decoding algorithms, each having its own trade-off between speed and accuracy. For surface codes, Minimum Weight Perfect Matching (MWPM)~\cite{higgott2021pymatching,higgott2023sparse} has a threshold higher than Union-Find (UF) ~\cite{Delfosse2021almostlineartime}, but is slower than the latter~\cite{delfosse2023choosedecoderfaulttolerantquantum,iolius2024decodingalgorithmssurfacecodes, maan2023testing}. Decoders are also usually specific to classes of codes~\cite{iolius2024decodingalgorithmssurfacecodes}: MWPM is usually applied to topological codes, whereas Belief Propagation + Ordered Statistics Decoding (BP+OSD) ~\cite{Panteleev_2021,roffe2020decoding} is very often used for QLDPC codes. A unification of graph based decoding algorithms is in~\cite{wu2024hypergraph}.

In general, BP does not have a threshold for the surface code and fails to decode most of the time~\cite{iolius2024decodingalgorithmssurfacecodes,roffe2020decoding}. Thus, a second stage decoder (e.g. OSD)~\cite{Panteleev_2021} does most of the decoding, and this fact increases the decoding runtime. Finding a (machine learning) method that can successfully decode QLPDC codes in a single stage has been an open problem.

\begin{figure}[!t]
    \centering
    \includegraphics[width=.49\textwidth]{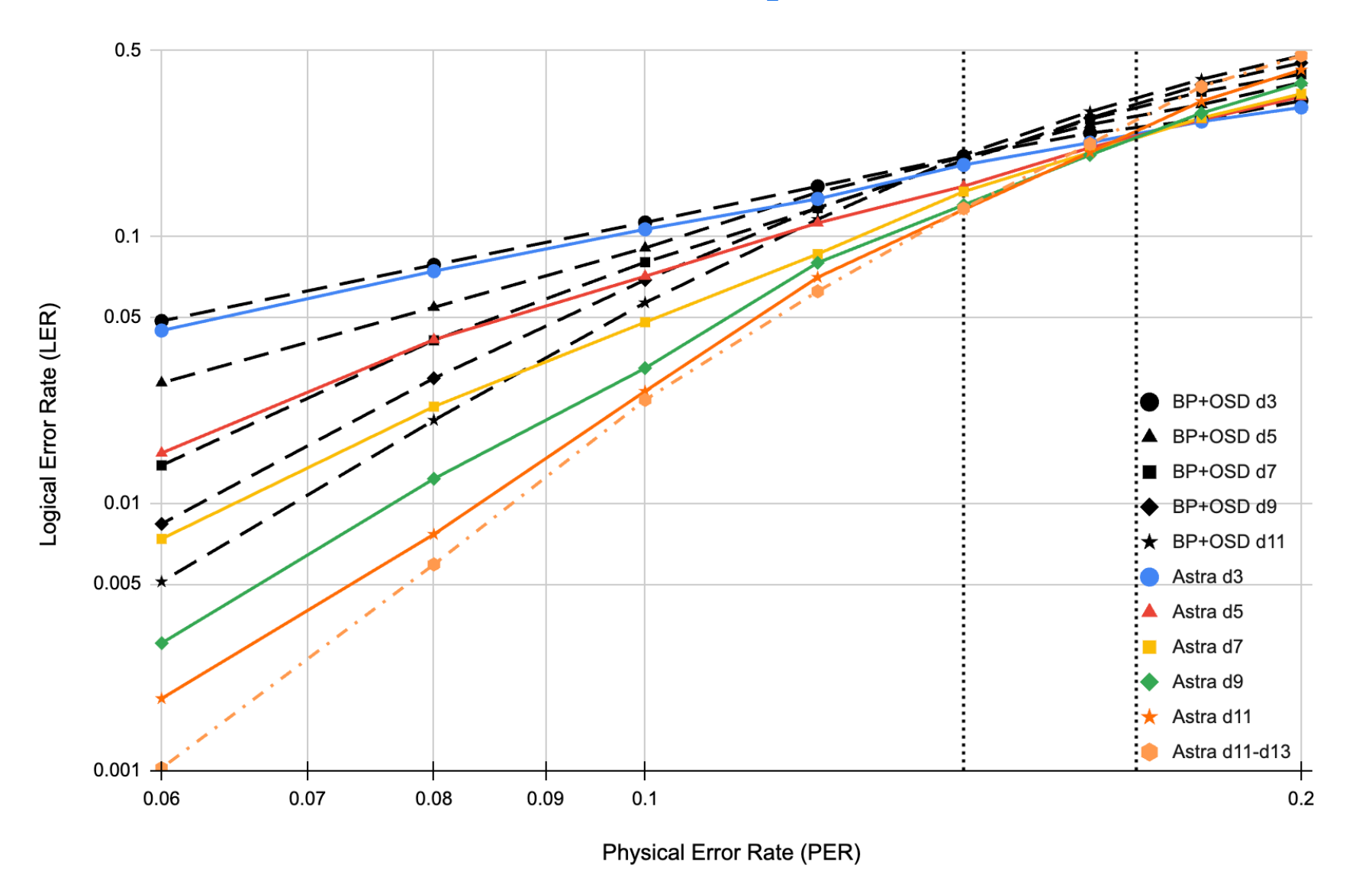}
    \caption{The Logical Error Rate (LER) of Astra vs BP+OSD under code capacity depolarizing noise. Our decoder has a threshold of $\sim 17\%$, and BP+OSD has a threshold of $\sim 14\%$.}
    \label{fig:ler_surface_bposd}
\end{figure}

\begin{figure}[!t]
    \centering
    \includegraphics[width=.49\textwidth]{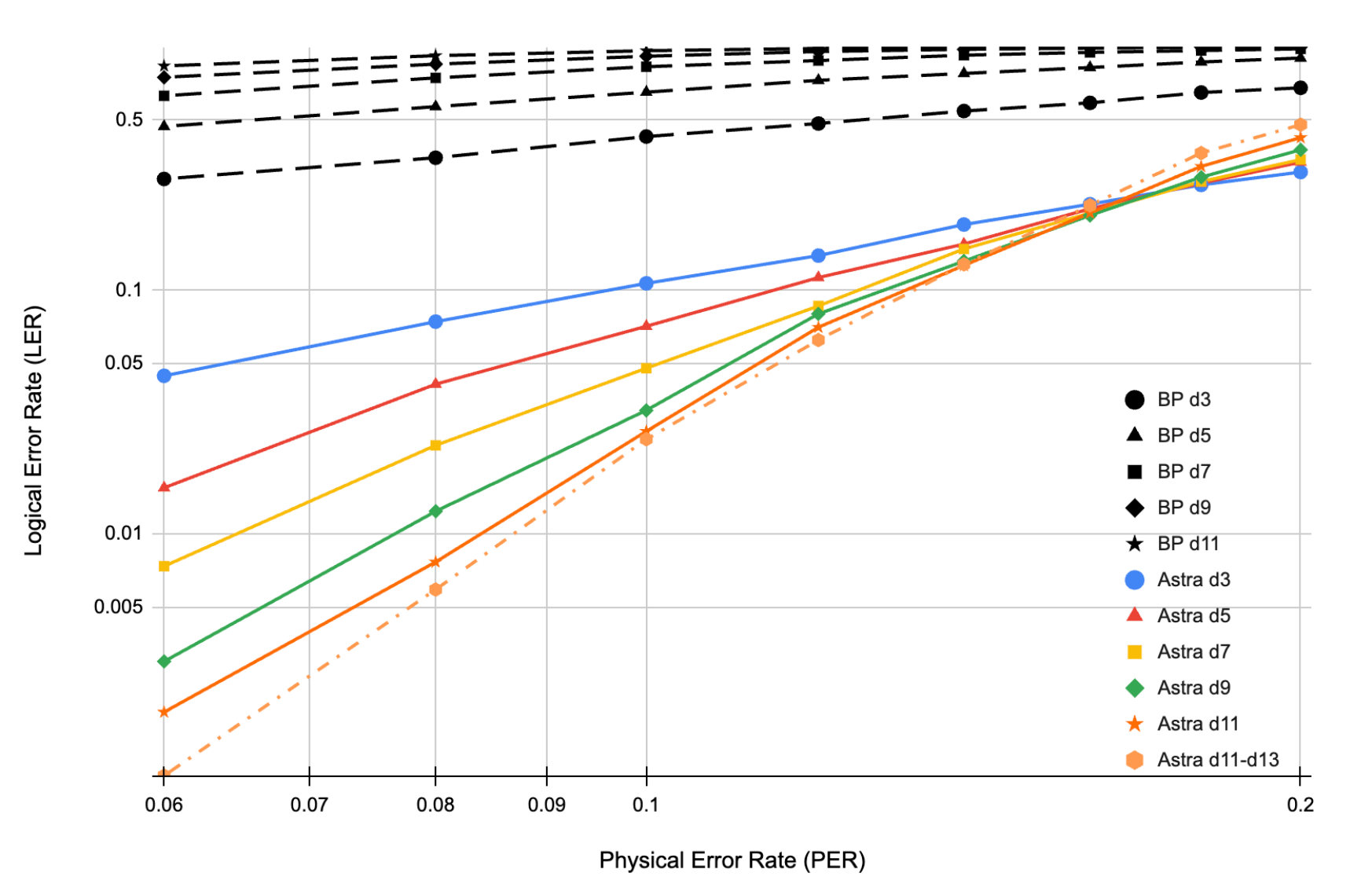}
    \caption{The Logical Error Rate (LER) for code capacity depolarising noise of Astra vs BP (no OSD). Our decoder outperforms BP, which does not have a threshold (i.e., the black dotted lines are not intersecting).}
    \label{fig:ler_surface_bp}
\end{figure}

To bridge this gap, we propose a novel machine learning decoder called Astra~\cite{maan2024scalable}. Our decoder is based on graph neural networks (GNN) which learn a message-passing algorithm for decoding. Astra can be used for arbitrary code distances, and outperforms BP+OSD (Fig.~\ref{fig:ler_surface_bposd}) and BP (Fig.~\ref{fig:ler_surface_bp}). 
Astra has the following features: 1) works on any family of codes that can be represented by Tanner graphs; 2) is capable of transfer learning; 3) has faster runtimes compared to BP+OSD while improving on accuracy.

In the following, we discuss how Astra improves the state-of-the-art of BP and machine learning decoders (Section~\ref{sec:related}). In Section~\ref{sec:methods} we discuss how our decoder is similar to solving Sudoku (Section~\ref{sec:sudoku}), we describe the loss function used for training (Section~\ref{sec:loss}) and the training procedure of Astra (Section~\ref{sec:training}). Section~\ref{sec:results} illustrates the performance of our decoder on surface and BB codes. We conclude and formulate future work in Section~\ref{sec:conclusion}. 

\subsection{Related Work}
\label{sec:related}

Recent work focused on improving the cost of the second stage decoder e.g. BP+CB (closed branch)~\cite{iolius2024closedbranchdecoderquantumldpc}, Ambiguity Clustering ~\cite{wolanski2024ambiguityclusteringaccurateefficient} and BP+LSD (Localised Statistics Decoding) ~\cite{hillmann2024localizedstatisticsdecodingparallel}). In contrast, Astra focuses on improving the first stage of decoding, such that the second stage can be completely avoided.

Independent of our work, the recent work of~\cite{ninkovic2024decodingquantumldpccodes} uses a similar GNN approach to decoding QLDPC codes. However, in contrast, Astra is capable of extrapolating both decoding and training (Section~\ref{sec:methods}), uses an improved loss function to overcome the challenges of learning in the presence of degenerate errors (Section~\ref{sec:loss}). Finally, Astra is more scalable and faster than BP+OSD and can be trained on commodity GPUs.

Machine learning decoders using neural networks (NN) have been proposed by~\cite{torlai2017neural,krastanov2017deep,varsamopoulos2017decoding,baireuther2018machine,chamberland2018deep,baireuther2019neural,andreasson2019quantum,varsamopoulos2019comparing,ni2020neural,wagner2020symmetries,sheth2020neural,varsamopoulos2020decoding,sweke2020reinforcement,meinerz2022scalable,ueno2022neo,chamberland2023techniques,overwater2022neural,gicev2023scalable,zhang2023scalable,egorov2023end,Liu_2019,miao2023quaternary,bausch2023learning,lange2023datadriven,varbanov2023neural,gong2023graph,wang2023transformerqec}. Each decoder comes with its own advantages and performance trade-offs, and the major scaling bottleneck is the training phase.

A majority of the NN decoders require some post-processing or a second stage decoder. Additionally, most of the NN decoders do not support transfer learning and need to be retrained in order to be used at different code distances. A model with transfer learning capabilities is ~\cite{wang2023transformerqec}, but their method uses MWPM as a second stage decoder. Neural belief propagation~\cite{Liu_2019} (NBP) performs better than the classical BP algorithm, but its transfer leaning capabilities were not investigated into detail.

A different approach to GNN decoding has been presented by~\cite{lange2023datadriven} where the model is learning the syndrome detection graphs. Their approach restricts the generalisations capabilities of the GNN, because the decoder learns some global functions rather than the low level connectivity of the data and syndrome qubits. Another recent GNN decoder is~\cite{gong2023graph} which works in tandem with BP and shows good performance for QLDPC codes, but the performance of the standalone GNN is not reported.

\subsection{Background}
\label{sec:back}

A code's threshold~\cite{gottesman2009introductionquantumerrorcorrection} is a function of the considered noise model (e.g. \emph{code capacity}, \emph{phenomenological}, or \emph{circuit level}) and the used decoder. In the code capacity noise model, Pauli errors on data qubits can happen with some probability $p$, while the syndrome (ancilla) qubits and their measurements are assumed perfect.

Under the circuit level depolarising noise model the surface code has a $1\%$ threshold ~\cite{wang2009thresholderrorratestoric,heim2016optimalcircuitleveldecodingsurface,Fowler_2012} --  one of the highest known so far. However, the surface code has a low encoding rate, because it encodes a single logical qubit. To overcome this limitation, high rate quantum LDPC codes were proposed. One of the recent LDPC codes are Bivariate Bicycle (BB)~\cite{bravyi2024highthreshold}. The BB code encoding 12 logical qubits requires only 288 physical qubits instead of 3000 physical qubits to give same level of error suppression as the surface code. The BB codes, under circuit level noise, have a high pseudo-threshold of $0.8\%$ - point when logical error rate goes below physical error rate.

\subsubsection{Belief Propagation and Message-Passing}

Belief propagation~\cite{bpbook} (BP) is a specific message-passing algorithm that has been extensively applied in the context of iterative decoding of classical Low-Density Parity-Check (LDPC) codes. More recently, BP has been adapted to the setting of QECC decoding (e.g.~\cite{babar2015fifteen}).

BP-based decoding algorithms for QECCs are trying to iteratively find the most probable assignment of values to the vertices of a Tanner graph, such that the assigned values agree with as many as possible of the syndrome measurements. The decoding algorithm terminates when a certain convergence criterion is met, or when a number of iterations exceeds a fixed threshold.

BP-based decoding algorithms are efficient in terms of computation time and memory requirements. However, BP is not always guaranteed to converge (see Fig.~\ref{fig:ler_surface_bp} where pure BP does not have a threshold) to the correct decoding solution. BP can suffer from issues like loopy behavior and local minima ~\cite{poulin2008iterative,roffe2020decoding,raveendran2021trapping,crest2024blindnesspropertyminsumdecoding, Panteleev_2021}.

Recently, Generalised-BP (GBP)~\cite{Old_2023} and Memory-BP (MBP)~\cite{Kuo_2022} have been proposed in order to achieve thresholds. GBP achieved a threshold comparable to BP+OSD and MWPM, but it has a higher complexity compared to BP. MBP achieves a code capacity threshold of $16\%$ for the surface code, but presents error floors at distances larger than 15, and the adaptive measures to overcome the floors are computationally costly.


\subsubsection{Graph Neural Networks}

GNNs are a class of neural networks used for learning representations of graph-structured data. GNNs are designed to operate on graphs directly, rather than on fixed-size, grid-like data such as images or sequences.

The intuition behind the functionality of GNNs is that these are iteratively updating node representations in a graph through message-passing. Each graph node aggregates information from its neighbors and updates its own representation based on the received information. The messages are defined by neural networks, and GNNs can be trained end-to-end using backpropagation to optimize a variety of objectives.

The specific architecture of a GNN can vary depending on the task and the nature of the graph data, but some common components include: a) node features; b) edge features; c) message network; d) graph convolutional layers; e) pooling layer; and f) output layer. Section~\ref{sec:methods} details on how some of these components are used during the training of Astra.

\section{Methods}
\label{sec:methods}

\begin{figure*}
    \includegraphics[width=0.8\textwidth]{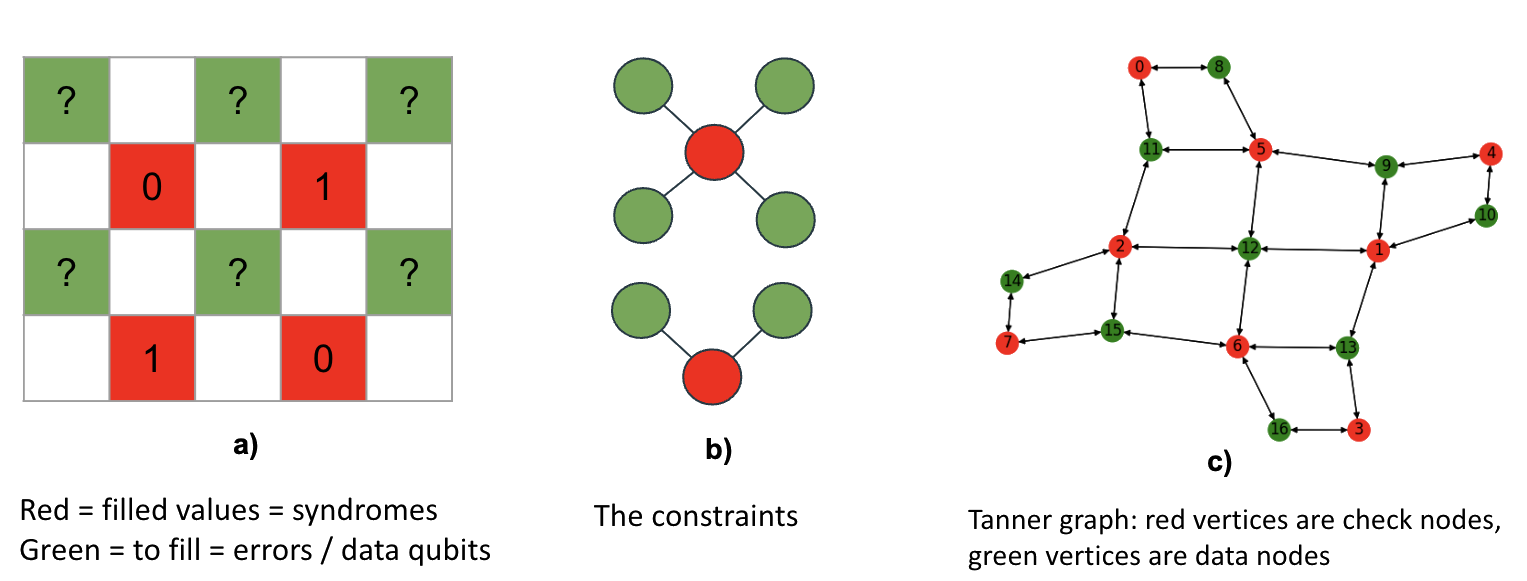}
    \label{fig:sudoku_decoding}
    \caption{Analogy between solving Sudoku and decoding (surface codes) by considering the Tanner graph as a graphical model of the constraints. a) the red cells are the already filled (syndrome qubit cells) and green cells (data qubits cells) need to be filled by the decoder; b) the value of the green cells depends on the values of the neighbouring red cells, such that in the rotated surface code there are weight-two and weight-four parity checks; c) the parities form the constraints which are encoded into the Tanner graph.}
\end{figure*}

We map the problem of decoding QLDPC codes to the problem of solving Sudoku using GNNs. Fig.~\ref{fig:sudoku_decoding} is an analogy of mapping initial filled cells from the Sudoku board to syndrome measurement values. Originally, GNNs were used by~\cite{palm2018recurrent} to solve $9\times 9$ Sudoku with at least 17 given values by encoding the constraints (e.g. Sudoku - a digit can exist only once in any row, column and $3 \times 3$ box) into a graph. 

Machine learning decoders have the potential to be generalized by implementing variants of \emph{transfer learning}\cite{pan2009survey,weiss2016survey}. To this end, we introduce the following definitions.

\begin{definition}[Interpolated decoding]
\label{ID}
We call interpolated decoding the action of decoding a distance $d'<d$ by using a decoder trained on distance $d$. (Example: Decode a distance 5 code by using a decoder trained for distance 11).
\end{definition}

Interpolated decoding is useful for improving the LER of the lower distance decoders. However, it is possible to use the pre-trained decoder in an approximate decoding mode.

\begin{definition}[Extrapolated decoding]
\label{ED}
We call extrapolated decoding the action of decoding a distance $d' > d$ by using a decoder trained on distance $d$. (Example: Decode a distance 15 code by using a decoder trained for distance 11).
\end{definition}

In the case of extrapolated decoding, the model has not seen training examples of code distance $d'$, and only decodes based on the knowledge obtained from smaller code distances.

The intuition behind extrapolated decoding is that a distance $d$ decoder will correct at most weight $(d-1)/2$ errors, but this behaviour will not change when extrapolating decoding to $d'>d$. Weight $d'>d$ errors are significantly less probable than weight $d$ errors. Therefore, extrapolated decoding leaves some errors uncorrected, yielding sub-optimal performance. Nevertheless, the sub-optimal extrapolated decoder can be combined with some second stage decoder such as OSD.

\begin{definition}[Extrapolated training]
\label{ET}
We call extrapolated training the action of training a decoder at distance $d' > d$ by starting from a decoder that has been pre-trained at distance $d$.
\end{definition}

We can speed up the training procedure by training a decoder for increasingly large distances. Instead of learning everything from scratch, one can reuse the previously accumulated knowledge. Therefore, once a decoder learns to correct $(d-1)/2$ errors, it preserves that knowledge irrespective of the distance and can correct at least as many errors even for $d' > d$.

\subsection{Solving Decoding Constraints using GNNs}
\label{sec:sudoku}

Decoding constraints are the checks on the data qubits. The constraints are encoded into the code's parity check matrix $H$. 

In the terminology of Sudoku, we will refer to data qubit nodes as \emph{error cells}, and syndrome qubit nodes as \emph{syndrome cells}. The latter are already filled by the syndrome data ($\sigma$) and given to the decoder.


The Sudoku decoding approach is to fill the error cells (the vector $e_{predicted}$) to satisfy the constraints corresponding to the parity checks touching the neighbouring error cells:

\begin{equation}
    He_{predicted} = \sigma\hspace{1mm}\mathbf{mod}\hspace{1mm}2
    \label{eq:he=s}
\end{equation}

The following section is based on ~\cite{palm2018recurrent} and describes how a GNN, in particular the Astra decoder, learns the decoding problem.

We use a GNN to learn the message-passing algorithm on the Tanner graph by: 1) unrolling the message-passing algorithm to a network; 2) learning the weights of this unrolled network.

The model learns three local functions: 1) how to send messages between nodes using a function \textbf{f}; 2) how to combine the messages and update the node states using a function \textbf{g}; 3) how to compute the final outputs from the node states using a function \textbf{r}. The functions are each learnt by different types of neural networks, as presented in the following.

At each iteration $t$ of the training procedure, the model is executing a sequence of steps. A hidden state vector $h^{t}_i$ is associated with each node $i$ of the graph. At the first step, the initial state of each node is $h^{0}_i = x_i$. In our case, $x_i$ is non zero for faulty syndromes and zero for error nodes.

\emph{Step 1.} Each node sends messages to its neighboring nodes. 
Let the message from node $i$ to node $j$ be $m_{i\rightarrow j}$.
\begin{equation}
    m^i_{i\rightarrow j} = \textbf{f}(h^{t-1}_i,h^{t-1}_j)
    \label{eq:mij}
\end{equation}
Here, the function \textbf{f} is a multi-layer perceptron (MLP) that learns the messages sent between the nodes.

\emph{Step 2.} Incoming messages are aggregated by
\begin{equation}
    m^t_{j} = \sum{m^t_{i\rightarrow j}}
    \label{eq:m_sum}
\end{equation}
where $N(j)$ is the neighbourhood of node $j$. In error correction terminology the neighbourhood is formed based on the parity checks encoded into $H$. from the parity checks.

\emph{Step 3.} Update the hidden node state
\begin{equation}
    h^t_{j} = \textbf{g}(h^{t-1}_i,x_j,m^{t}_j)
    \label{eq:hupdate}
\end{equation}
where \textbf{g} is another neural network which learns the update rule that depends on the previous node state, input state and the aggregated messages. We use a gated-recurrent unit (GRU) to learn the $g$ function.

4. Finally, \textbf{r} is obtained by learning the distribution of the hidden node states $h^t_i$ via another MLP. 
\begin{equation}
    o^t_{i} = \textbf{r}(h^{t}_i)
    \label{eq:outprob}
\end{equation}

In the case of Astra, the output probabilities correspond to the probabilities attached to the error nodes. The output probabilities are then included in the loss function
(see next section), which enables the supervised training of the GNN model.

In a nutshell, instead of decoding using a monolithic, large NN which predicts errors given the syndromes, Astra is converting the decoding problem into the one of learning three local functions \textbf{f, g} and \textbf{r}. The functions work according to the local constraints from the Tanner graph.

The degree of the nodes does not change with increasing QECC distance. Once learnt, the functions can be reused for interpolated and extrapolated decoding, as well as extrapolated training. Section~\ref{sec:results} illustrates the capabilities of the model and its training procedure.

\subsection{Loss Function}
\label{sec:loss}

The large \emph{degeneracy} (i.e. many errors generate the same syndrome) of the QECCs is a serious challenge when training machine learning models. The original Sudoku approach presented by~\cite{palm2018recurrent} does not include degenerate solutions: there is a unique Sudoku solution if there are at least 17 filled cells in the $9 \times 9$ board.

To overcome the challenges of training in the presence of degeneracy, we use a loss function which consists of two parts: 1) the cross entropy loss between the output probability distribution and the actual one; 2) the loss function motivated by NBP~\cite{Liu_2019}. The latter ensures that the predicted correction, when applied to the actual error, commutes with all the stabilisers and the logical operators:
\begin{equation}
    H^{\bot}e_{total} = 0 \hspace{1mm}\mathbf{mod}\hspace{1mm}2,
    \label{eq:heperp=0}
\end{equation}
where,
\begin{equation}
    e_{total} = \left(e_{actual} + e_{predicted}\right)\%2 
\end{equation}

and $H^{\bot}$ is the orthogonal complement of the parity check matrix~\cite{Liu_2019}.

Our loss function helps Astra to learn the degenerate error-syndrome pairs by simultaneously minimising  Eq.~\ref{eq:heperp=0} and the cross entropy loss. To minimise Eq.~\ref{eq:heperp=0} in a machine leaning setting, the $\mathbf{mod}\hspace{1mm}2$ is replaced with the differentiable function $|sin(\pi*x/2)|$, where $x$ is the left hand side of Eq.~\ref{eq:heperp=0}. Note that in Eq.~\ref{eq:heperp=0}, both the parity check matrix $H$ and $e_{total}$ represent one error type and the loss is minimised for X and Z errors separately -- this is un-correlated decoding. 

\begin{figure*}
    \centering
    \includegraphics[width=.45\textwidth]{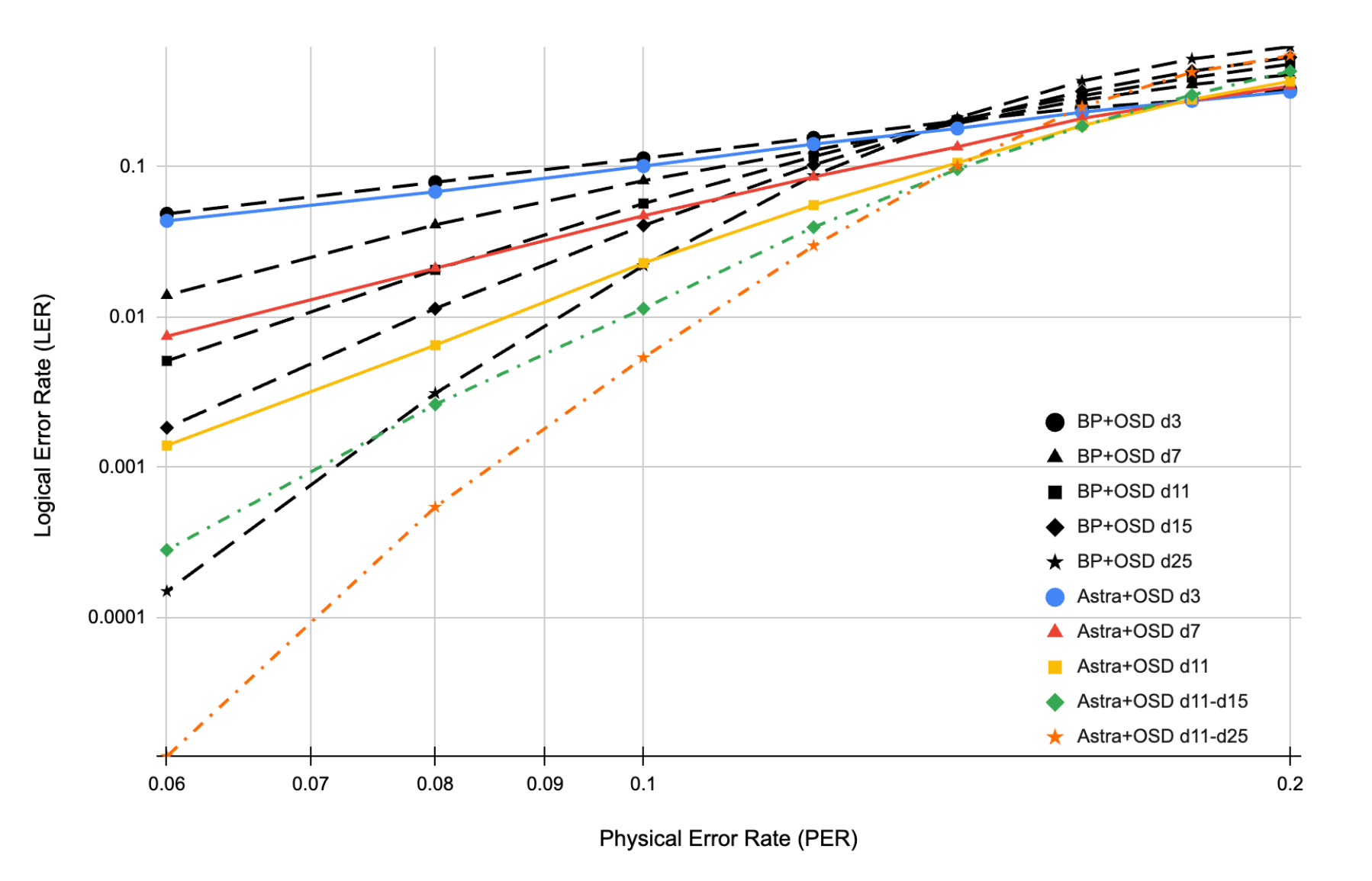}
    \includegraphics[width=.45\textwidth]{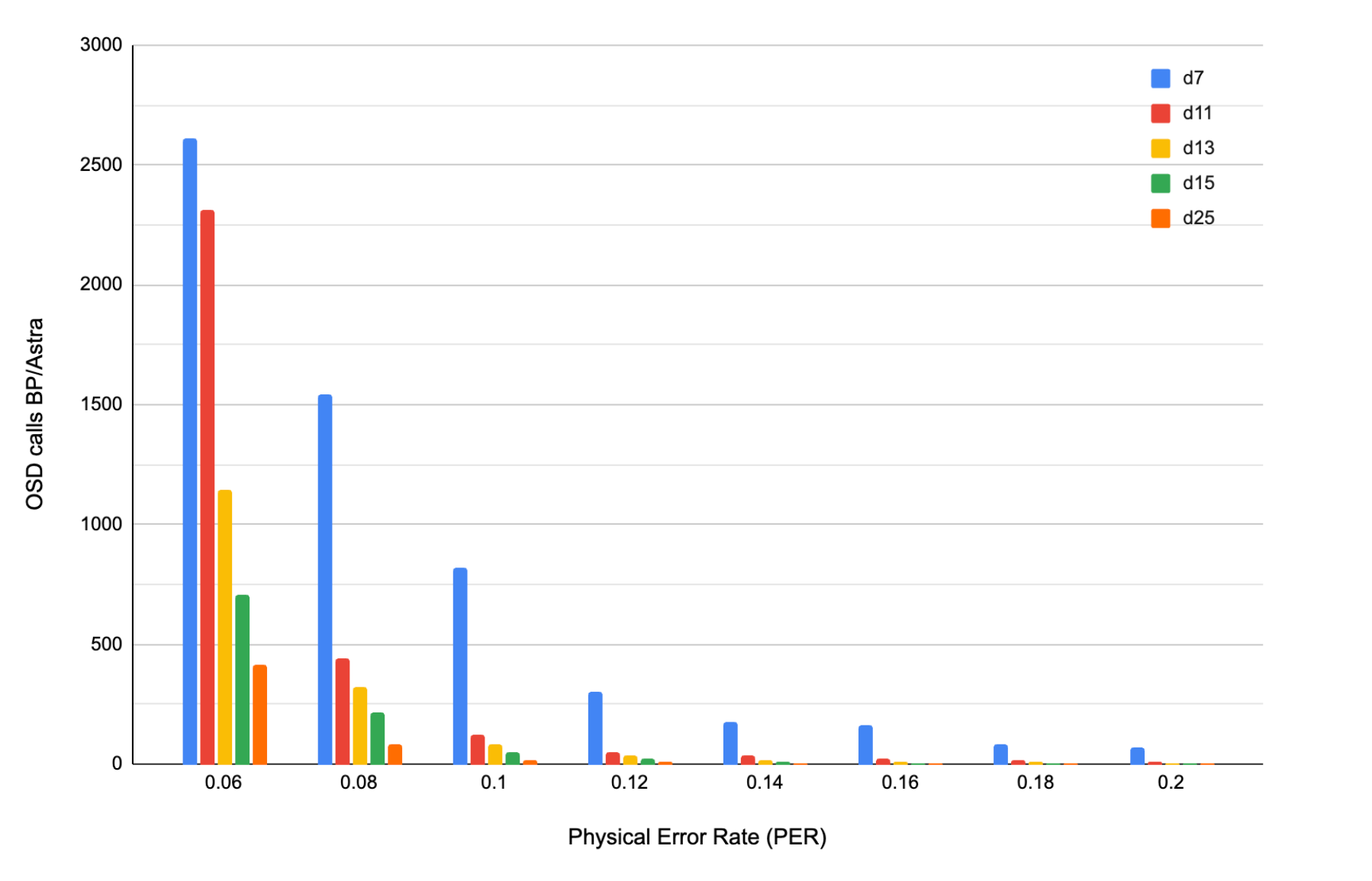}
    
    \caption{Decoding surface codes with Astra+OSD vs BP+OSD by using OSD0 in the second stage. a) Astra+OSD achieves orders of magnitude better LER than BP+OSD and requires fewer OSD calls; b) Speedups of Astra+OSD vs BP+OSD are obtained because Astra converges more often than BP and, consequently, the OSD stage is called significantly fewer times. This holds even when performing extrapolated decoding  with the d11 decoder e.g. for distance 25, at 0.06 error rate, Astra+OSD is 400x faster than BP+OSD.}
    \label{fig:ler_sc_osd_calls}
\end{figure*}

\begin{table*}    
    \centering
    \caption{Astra hyperparameter values achieving lowest LER.}
    \label{tab:training_log}
    \begin{tabular}{c|c|c|c|c|c|c|c|c|c|c|c}
    Code    & d & Num. of Iterations  & Node features  & Edge features  &  Msgnet size & Training set  &Test set\\
    \hline
    Surface & 3 & 30        & 500       & 500       & 512      & $10^5 @ 0.15$     & $10^4 @ 0.05$\\
         -  & 5 & 30        & 500       & 500       & 512      & $2*10^5 @ 0.15$   & $10^4 @ 0.05$\\
          - & 7 & 45        & 500       & 500       & 512      & $4*10^5 @ 0.15$   & $10^4 @ 0.05$\\
          - & 9 & 100       & 500       & 500       & 512      & $5*10^5 @ 0.15$   & $10^4 @ 0.05$\\
     -      & 11& 100       & 500       & 500       & 512      & $5*10^5 @ 0.15$   & $10^4 @ 0.05$\\
    \hline
    BB      & 6 & 30        & 50        & 50        & 128      & $2*10^4 @ 0.15$   & $10^3 @ 0.05$\\
        -   & 12& 30        & 50        & 50        & 128      & $2*10^4 @ 0.18$   & $10^3 @ 0.08$\\
        -   & 18& 45        & 50        & 50        & 128      & $10^5 @ 0.15$     & $5*10^3 @ 0.1$\\
    \end{tabular}
\end{table*}

\subsection{Correlated Decoding}

We benchmark Astra using the code capacity depolarising noise model (Section~\ref{sec:back}), and there are three possible Pauli errors: X, Z and Y (decomposed into X and Z errors). Most of the decoders do independent (un-correlated) decoding for X and Z channels, but this technique fails to decode correlated Y errors and results in sub-optimal decoding performance.

To learn the message-passing decoding of correlated errors, we use 1-bit hot encoding for errors where X errors are represented by 1, Z errors are 2 and Y errors are 3. This helps the decoder to learn the relationship between different syndrome qubit flips generated by different types of errors.

\subsection{Training}
\label{sec:training}

Astra is currently implemented and trained using \emph{PyTorch}. During training, we use disjoint training and test datasets composed of error-syndrome tuples.  We performed hyperparameter search using \emph{optuna}~\cite{akiba2019optuna}. The hyperparameters of the successful trainings are tabulated in Table~\ref{tab:training_log}. Additionally, we used a learning rate of $1e-4$, a dropout rate of $0.05$ in the message net layer. The training is stopped when: 1) the LER of the decoder is substantially lower than the physical error rate for the test set, and 2) the loss over the test set converges to zero.

We test Astra on a code capacity depolarizing noise model where we assume noise only on data qubits and perfect syndrome qubits. The training and test datasets are composed of error-syndrome tuples. Data is generated using the \emph{panqec}~\cite{Huang_2023} and the \emph{ldpc}~\cite{roffe2020decoding,Roffe_LDPC_Python_tools_2022} software packages.

We train a single model for a given code distance and multiple physical error rates $p$. To ensure the model sees sufficient error patterns we first sample random $p$ values from a uniform distribution $U(0, p_{max})$ where $p_{max}$ is typically chosen to be around the theoretical maximum threshold of the QECC. We generate error-syndrome tuples for each value of $p$.

\begin{figure*}
    \centering
    \includegraphics[width=.45\textwidth]{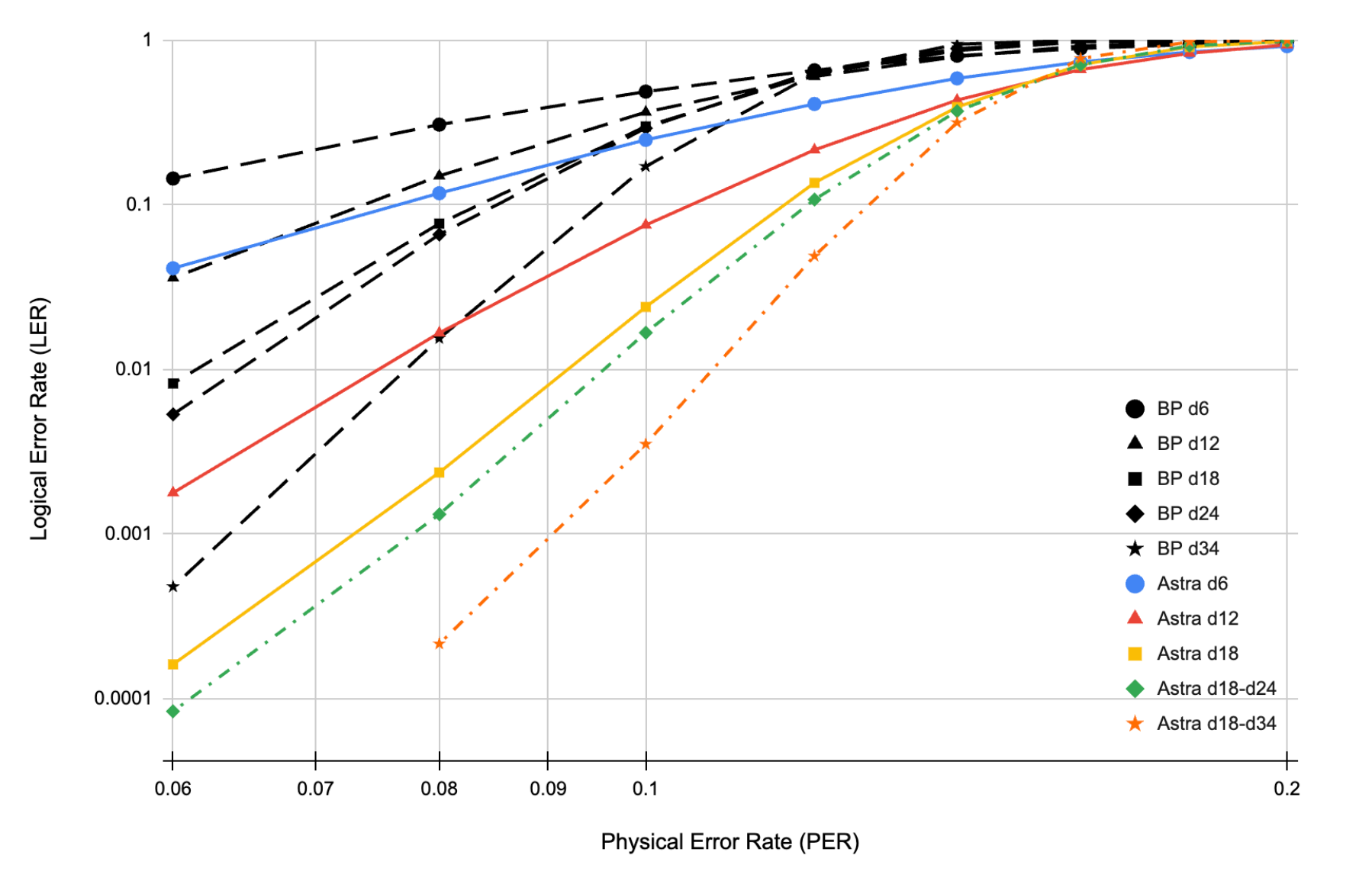}
    \includegraphics[width=.45\textwidth]{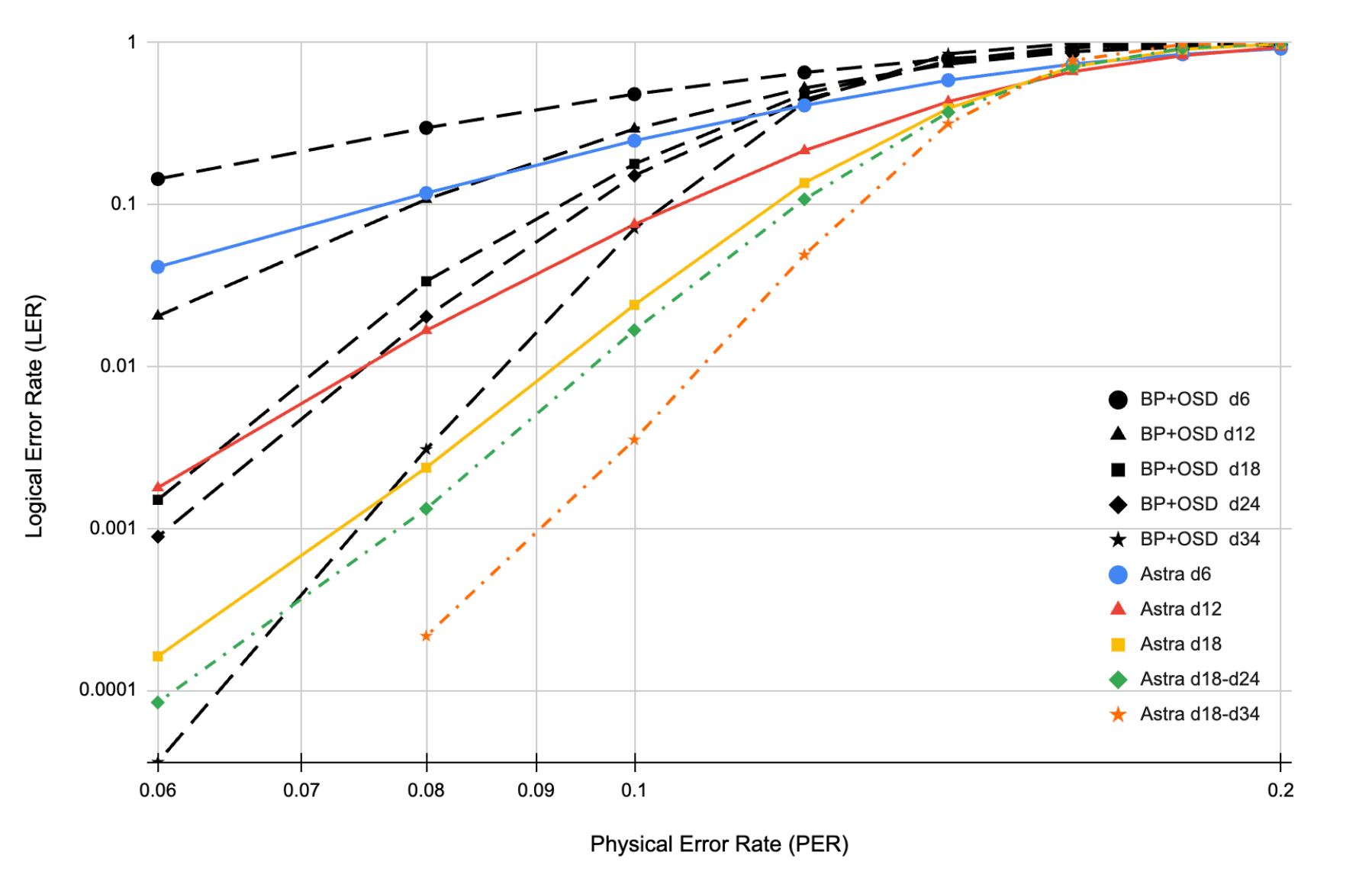}
    
    \caption{Decoding BB codes with Astra compared to BP and BP+OSD. a) The LER of Astra is significantly lower compared to pure BP; b) The LER of Astra compared to BP+OSD.}
    \label{fig:ler_bb} 
\end{figure*}

\begin{figure*}
    \centering
    \includegraphics[width=.45\textwidth]{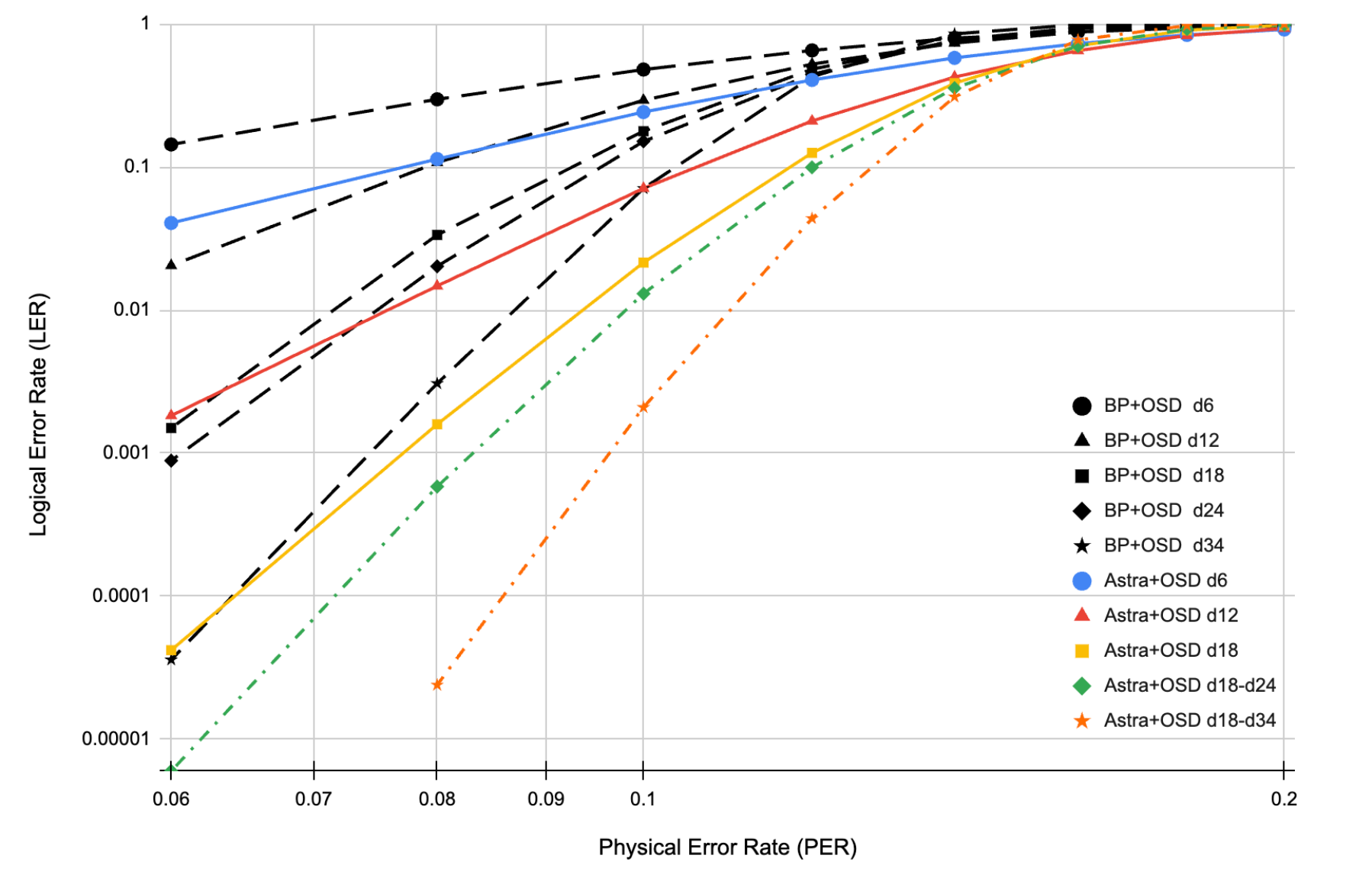}
    \includegraphics[width=.45\textwidth]{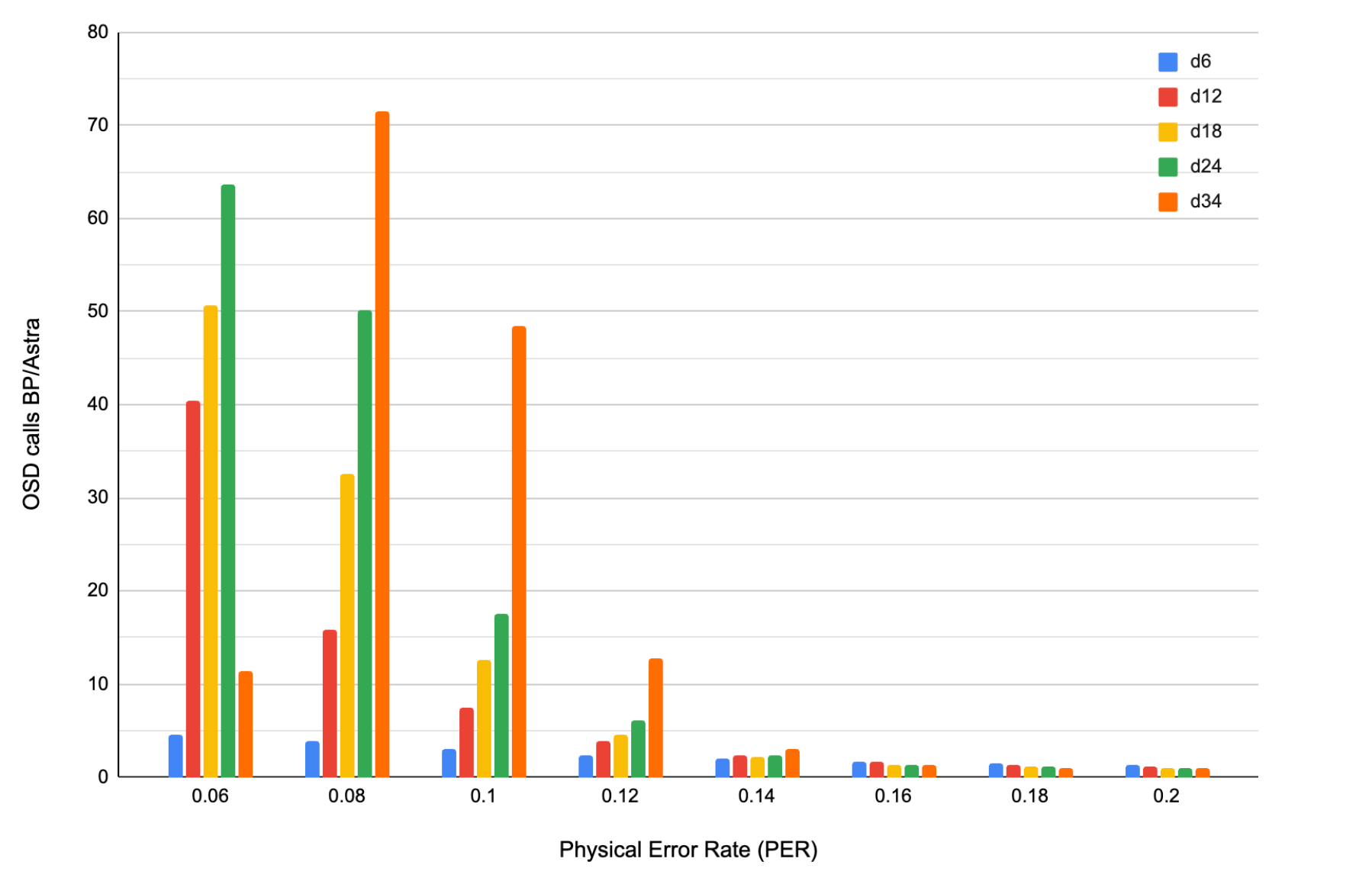}
    \caption{Decoding BB codes with Astra+OSD compared to BP and BP+OSD. a) LER of Astra+OSD vs BP+OSD; b) Speed-up of Astra+OSD vs BP+OSD, Astra+OSD is $\sim 50x$ faster than BP+OSD for larger codes at low errors rates. The speedups persists even for the extrapolated decoding case of distance 24 and 34 using distance 18 GNN decoder.}
    \label{fig:ler_bb2} 
\end{figure*}

Our approach ensures that the GNN model works for a given distance irrespective of the error rate. To further boost the learning quality, we replace the training and test sets before each training epoch. The test data is generated for a fixed $p$ and is used to calculate the test loss and to monitor training progress.

We also executed extrapolated training of large distances for surface code and BB code. We trained d11 surface code starting the training from the d9 trained model. Similarly, we trained BB d12 decoders starting from d6, and BB d18 starting from d12. We observed massive speed-ups of the training procedure without sacrificing the LER of the decoders.

All the training for BB code up to distance 18 and for the surface code up to d7 was executed on a \emph{NVIDIA RTX-3080} GPU, and took at most 24h.

\section{Results}
\label{sec:results}

Herein we present the results of decoding surface codes and Bivariate Bicycle quantum LDPC codes~\cite{bravyi2024highthreshold}.

The GNN has (approximately) learned a message-passing algorithm that satisfies the decoding constraints expressed as a Tanner graph. The model can be used for extrapolated decoding and, if necessary, for extrapolated training. We will illustrate the decoding performance in the context of Definitions~\ref{ID} and~\ref{ED}.

Decoding is performed by running Astra for a specific number of iterations. This is similar to how BP is used during syndrome decoding~\cite{roffe2020decoding}. We can adapt the maximum number of iterations. In our experience, adapting the maximum number of iterations improves the LER.

We observe that Astra has higher LERs compared to BP+OSD. We define the speed-up between Astra+OSD and BP+OSD as the ratio between how many times BP is failing and the number of times Astra is failing.

Fig.~\ref{fig:ler_sc_osd_calls} shows that when using Astra for surface code, the second stage decoder is called orders of magnitude fewer times, resulting in orders of magnitude speedup. For the distances we train (up to d11), we get more than 2000X speedup at very low error rates compared to BP+OSD. For the extrapolated decoding case, decoding d25 using d11 yields speed-ups of 400x at low error rates. We observe up to 40x speed-ups for the BB code (see Fig.~\ref{fig:ler_bb2}).

\subsection{Decoding Surface Codes}

Figs. ~\ref{fig:ler_surface_bposd} and~\ref{fig:ler_surface_bp} show the simulation results of the GNN decoder for surface code. Fig.~\ref{fig:ler_surface_bposd} shows that standalone Astra, without using any post-processing, is outperforming BP+OSD for code capacity depolarising noise model. For example, the Astra d9 outperforms BP+OSD's d11. This might indicate that Astra learned to correct correlated errors better than BP+OSD.

All surface code decoding results were obtained using a model trained on d11 -- a case of interpolation decoding. Additionally, we perform extrapolated decoding by starting from the d11 decoder. The d13 curve is generated by decoding with decoder trained on d11 and the decoder corrects weight 5 error for d13 too.

\subsection{Decoding Bivariate Bicycle Codes}

Figure~\ref{fig:ler_bb} illustrates Astra's decoding performance of Bivariate Bicycle codes under code capacity depolarising noise. Astra(+OSD) achieves orders of magnitude lower LERs for BB codes compared to BP(+OSD). Moreover, Astra shows impressive extrapolated decoding capabilities when decoding the $[360,12,24]$ and $[756,16,34]$ codes with GNN trained on $[288,12,18]$ code. Astra's extrapolated behaviour is better than BP+OSD.

\section{Discussion}

For all the simulation data with the BP decoder we used the serial scheduling, a maximum of 100 iterations, the minimum sum algorithm and a scaling factor of 0.75 for BB codes and a scaling factor of 1.0 for surface codes. For Astra we used the flooding scheduling i.e. sending all the messages at once. In general, BP has better decoding performance when using serial scheduling rather than flooding. 
One can, in principle, implement different scheduling methods with GNN too, be it serial, parallel or random. We did not experiment with these but we expect that changing the scheduling might help the Astra decoder even better like in the case of BP. 



\section{Conclusion}
\label{sec:conclusion}

We introduced Astra, a novel, scalable graph neural network decoder that is trained on the Tanner graph of QLDPC codes, such as the surface and BB codes. Our decoder does not require post-processing in order to outperform BP+OSD. When using OSD, Astra+OSD is significantly faster than BP+OSD. Our decoder can be trained easily on commodity GPUs and exhibits features of transfer learning. We are using Astra to decode distances which are higher than the ones used during training, and we are still outperforming BP+OSD. Future work will focus on extending the current results to circuit-level noise.

\begin{acknowledgments}
We thank Ioana Moflic and Huyen Do for their feedback on a first version of this manuscript. This research was developed in part with funding from the Defense Advanced Research Projects Agency [under the Quantum Benchmarking (QB) program under award no. HR00112230006 and HR001121S0026 contracts], and was supported by the QuantERA grant EQUIP through the Academy of Finland, decision number 352188. The views, opinions and/or findings expressed are those of the author(s) and should not be interpreted as representing the official views or policies of the Department of Defense or the U.S. Government. We acknowledge the computational resources provided by the Aalto Science-IT project.

\end{acknowledgments}

\bibliography{__main}

\end{document}